\documentclass[reprint,amsmath,amssymb,aps]{revtex4-2}

\usepackage{graphicx}
\usepackage[colorlinks,citecolor=blue]{hyperref}
\usepackage{float}
\usepackage{booktabs,tabularx}
\usepackage{tikz}
\usetikzlibrary{calc}
\usetikzlibrary{arrows.meta}
\usetikzlibrary{decorations.pathmorphing}
\usetikzlibrary{shapes.symbols}
\usetikzlibrary{positioning}
\usetikzlibrary{fit}

\usepackage[utf8]{inputenc} 
\usepackage[T1]{fontenc}    

\usepackage{mathtools}
\usepackage{url}

\DeclarePairedDelimiter\ket{\lvert}{\rangle}
\DeclarePairedDelimiterX\braket[2]{\langle}{\rangle}{#1\,\delimsize\vert\,\mathopen{}#2}

\begin{document}

\title{High-Fidelity Transmon Reset with a Multimode Acoustic Resonator}

\author{Andraž Omahen$^{1,2}$}
\email{aomahen@phys.ethz.ch}
\author{Simon Storz$^{1,2}$}
\author{Igor Kladarić$^{1,2}$}
\author{Yiwen Chu$^{1,2}$}
\affiliation{$^{1}$Department of Physics, ETH Z\"{u}rich, 8093 Z\"{u}rich, Switzerland}
\affiliation{$^{2}$Quantum Center, ETH Z\"{u}rich, 8093 Z\"{u}rich, Switzerland}

\date{\today}

\begin{abstract}
Achieving sufficiently low residual excited-state populations remains a key challenge in superconducting quantum circuits, particularly for protocols operating close to noise limits or requiring repeated qubit initialization.
Existing protocols primarily address this challenge through sophisticated control, engineered dissipation, or feedback mechanisms. Here, we demonstrate an alternative approach in which a superconducting qubit is reset using a physically distinct, intrinsically colder phononic bath. Specifically, we interface a transmon with a high-overtone bulk acoustic resonator (HBAR), enabling cooling of the qubit into GHz-frequency modes. Using this approach, we achieve a mean residual excited-state population of the qubit below $10^{-4}$ and an upper error bar below $3 \times 10^{-4}$ with 95\% confidence, representing an improvement of one to two orders of magnitude compared to typical reset schemes. These results highlight the potential of phononic baths as a resource for high-fidelity qubit initialization in superconducting circuits.

\end{abstract}

\maketitle

Initializing quantum information processing systems in the ground state of their computational subspace is a critical requirement for quantum computing and sensing. Over the last two decades, superconducting circuits have matured to one of the most promising platforms for large-scale quantum information processing tasks \cite{Kim2023, Acharya2025, Gyawali2025}, and they have emerged as serious contenders for sensing applications \cite{Braggio2025,DeDominicis2025, Lescanne2020, Li2025sensing}. In order to reach the required quantum regime where $k_b T \ll \hbar \omega$, superconducting circuits are passively cooled in dilution refrigerators. However, since the circuits are open electromagnetic systems that couple to multiple external heating mechanisms, typical residual excited state populations still regularly reach the percent level \cite{Yan2016, Yeh2017,Yan2018,Wang2019}. For high-fidelity operations, long sequences, or particularly sensitive and noise-limited protocols, this level of initialization is often insufficient. In fact, imperfect ground state initialization has been identified as one of the limiting factors in experiments requiring repeated, high-fidelity state preparation, including quantum error-correction for applications in quantum computing \cite{Krinner2022}, or single-photon \cite{Inomata2016} or dark matter \cite{Braggio2025} detection for use cases in quantum sensing.

For this reason, in addition to hardware improvements \cite{Yeh2017,Wang2019}, a variety of active reset schemes have been developed for better qubit initialization. These protocols trade off speed, fidelity, and experimental complexity. Simple protocols rely on existing techniques such as projective measurements \cite{Riste2012} or the simultaneous driving of qubit and cavity transitions \cite{Geerlings2013}, and are widely used in experiments with superconducting circuits. Faster protocols typically require carefully engineered microwave resonators with complex pulse calibrations \cite{Magnard2018, Egger2018}, additional on-chip elements such as diplexers or couplers \cite{Ding2025, Zhou2021}, or active feedback using an FPGA \cite{Salathe2018}. Protocols focusing on particularly high fidelities rely on external microwave switches \cite{Han2023} or careful bath engineering \cite{Aamir2025}.

Despite their differing implementations, these approaches operate within the same electromagnetic environment. Here, we instead couple a transmon qubit to a physically different, phononic bath using a high-overtone bulk acoustic resonator (HBAR) \cite{Chu2017, Chu2018}.
 These acoustic resonators provide intrinsically colder effective bath temperatures, as they couple to the environment through different channels than their microwave counterparts. As a result, they are substantially less affected by the dominant heating mechanisms in circuit QED systems, including coupling to electromagnetic noise \cite{Rower2023}, high-energy radiation \cite{Kerschbaum2026}, quasiparticles \cite{McEwen2022}, and photons from the measurement process \cite{Sank2026}. 
At dilution-refrigerator temperatures, the GHz-frequency acoustic modes of an HBAR therefore exhibit significantly lower thermal occupation than microwave modes, with recent measurements reaching steady-state populations below $10^{-4}$ \cite{Omahen2025}.

In addition, we exploit the intrinsic and readily accessible multimode nature of HBAR devices, which provides a hardware-efficient mechanism for repeated entropy extraction without introducing additional dissipation channels or control hardware. This allows us to engineer the reset of a superconducting circuit qubit to a residual excited-state population on the order of $10^{-4}$, a factor 5-10 lower than even the highest fidelity schemes so far \cite{Aamir2025, Zhou2021}, while keeping a moderate duration comparable to many existing protocols, and maintaining modest experimental complexity. The only additional component required is the HBAR, and the relatively simple protocol does not involve any active feedback.

\section*{Reset Protocol}

\begin{figure}
  \centering
  \includegraphics[width=1.0\columnwidth]{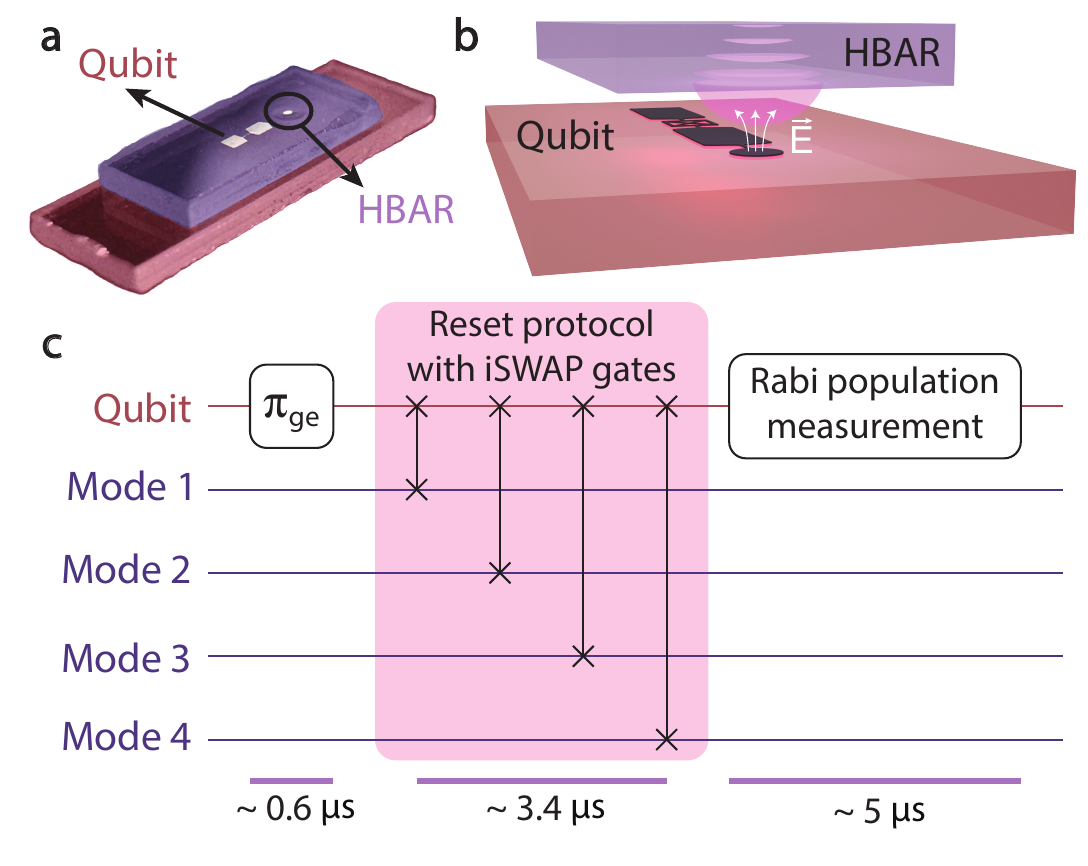}
  \caption{\textbf{Experimental System and Protocol.} 
  \textbf{a}, Photograph of an HBAR (top chip) bonded to a superconducting circuit qubit (bottom chip).
  \textbf{b}, Schematic of the device: HBAR (top) coupling to a transmon (bottom) through a piezoelectric material (dome).
  \textbf{c}, Experimental protocol for the qubit reset and characterization procedure.}
  \label{fig:protocol}
\end{figure}

Our device (Fig.~\ref{fig:protocol}a) is assembled using a flip-chip bonding technique, where a superconducting transmon qubit (bottom chip in Fig.~\ref{fig:protocol}a) is coupled to an HBAR (top layer). 
The coupling between the transmon electric field and the acoustic modes is mediated by a piezoelectric dome. The HBAR acts as a Fabry-Perot cavity for phonons, confining many acoustic modes with Gaussian beam profiles that are separated in frequency by a free spectral range of about 12.6~MHz. The combined transmon-HBAR system is approximated by the Jaynes-Cummings Hamiltonian
\begin{equation}
   H / \hbar = \omega_q q^\dagger q - \frac{\alpha}{2} {q^\dagger}^2 q^2  + \sum_i \omega_i a_i^\dagger a_i + \sum_i g_i (a_i q^\dagger + a_i^\dagger q),
\end{equation}


where $\omega_q$ stands for the qubit frequency, $\alpha$ for the anharmonicity, and $q$ ($q^\dagger$) for the qubit annihilation (creation) operator. $\omega_i$ is the frequency of the acoustic mode with mode number $i$ and the corresponding annihilation and creation operators $\hat{a}_i$ and $\hat{a}_i^{\dagger}$. Applying an off-resonant Stark shift drive to the qubit allows us to tune its frequency.
When tuning the qubit on resonance with phonon mode $i$, the two systems coherently exchange excitations $\left| e \right> \left| 0 \right>_i \leftrightarrow  \left| g \right> \left| 1 \right>_i $ at rate $g_i$. 
An iSWAP gate between qubit and phonon mode is achieved by having the two systems resonantly interact for time $t_{\mathrm{iSWAP}} = \pi / (2 g_i) $ \cite{Chu2017, Chu2018}. In our reset protocol, we employ this process to extract entropy from the hot qubit to the cold phonon modes by repeatedly applying iSWAP gates with different phonon modes (Fig.~\ref{fig:protocol}c).






\section*{Performance Characterization}
To characterize the reset performance, we prepare the qubit in the excited state $\ket{e}$ (Fig.~\ref{fig:protocol}c), followed by the reset protocol - a series of iSWAP gates between the qubit and different acoustic modes - and finally a measurement of the qubit state (see Appendix \ref{app:full_protocol} for full details). 

The extraction of the residual excited state population of the qubit is performed using a two-point Rabi population measurement (RPM) \cite{Geerlings2013, Jin2015, Omahen2025}. RPM measures the amplitude of Rabi oscillations between the first excited state $\ket{e}$ and second excited state $\ket{f}$ of the qubit, which results in the signal amplitude. This is then repeated while adding an initial $\pi_{ge}$-pulse that swaps the population of the ground state $\ket{g}$ and the state $\ket{e}$, which results in the reference amplitude. The measurement outcome is RPM contrast, given by the ratio of the signal and reference amplitudes. RPM contrast uncertainty is dominated by the much smaller signal amplitude, while the uncertainty of the reference amplitude contributes only weakly, as followed from Gaussian error propagation.
This method assumes that the population of the second excited state is much smaller than the population of the first excited state. We discuss this aspect in Appendix~\ref{app:f_level}. 

\begin{figure}
  \centering
  \includegraphics[width=1.0\columnwidth]{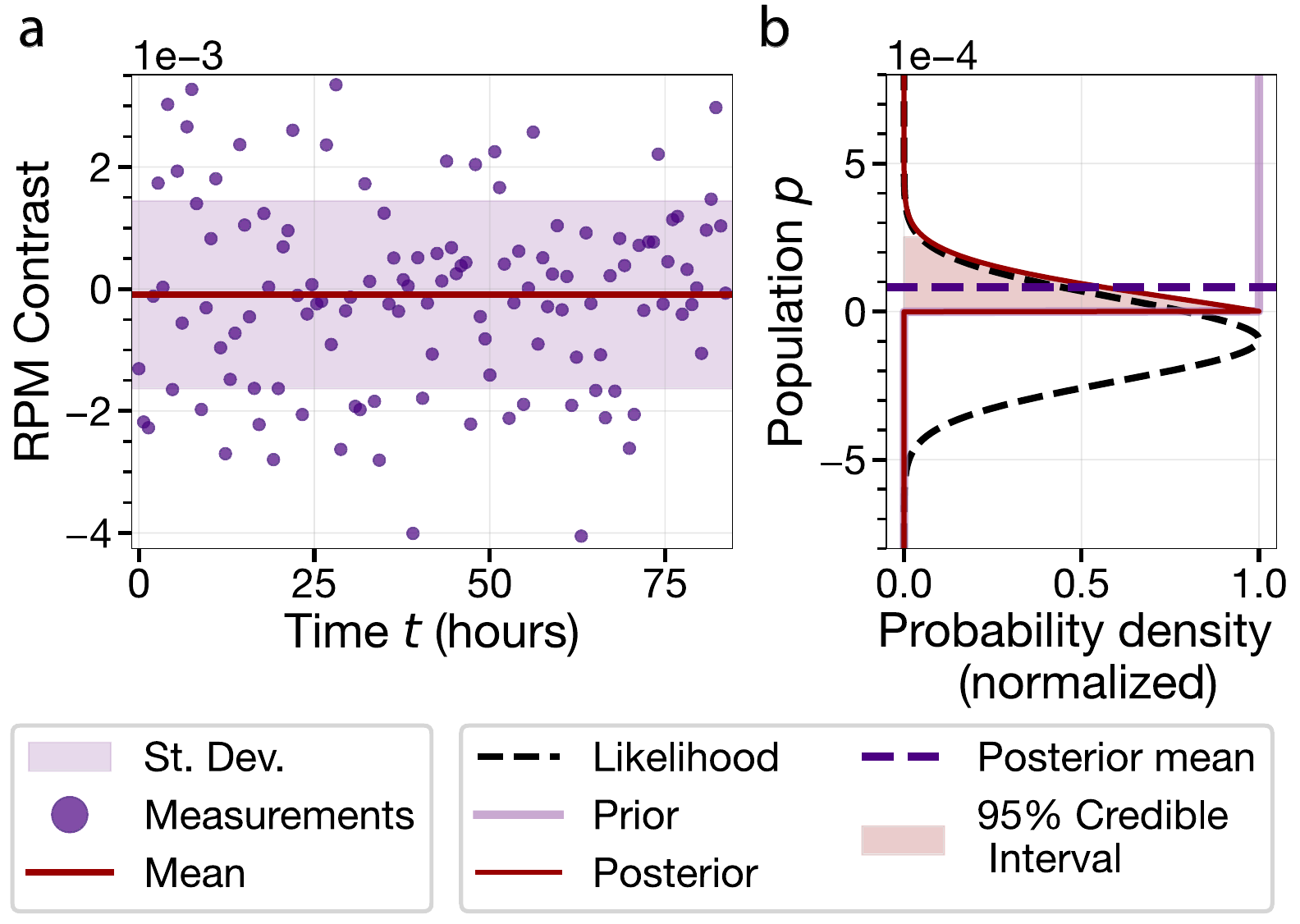}

  \caption{\textbf{Residual population measurement and Bayesian analysis.} 
  \textbf{a}, Measured RPM contrast for a single reset characterization measurement with four SWAP gates, acquired over the course of about 3.5 days. 
  \textbf{b}, Baysian analysis of the RPM data for the extraction of the residual populations. The purple dashed line represents the mean of the posterior distribution, while the red region denotes the $95 \% $ confidence interval. The probability densities on the x-axis are normalized to reach the maximum value of 1. 
  }
  \label{fig:RPM-data}
\end{figure}

An example of a reset characterization using four iSWAP gates is shown in Fig.~\ref{fig:RPM-data}a. Because of the low residual populations and limitations in the signal-to-noise ratio of the measurement process, the data is acquired over the course of 84 hours in order to reduce the statistical uncertainty. Each point was averaged $5 \times 10^5$ times, and the resonance amplitude and optimal swap duration $t_{\mathrm{iSWAP}}$ were recalibrated in between each point. Averaging all the data yields a mean value of RPM contrast $\mu_4 = - 0.95 \times 10^{-4}$ and a standard deviation of the mean $\sigma_4 = 1.39 \times 10^{-4} $. 

We then use Bayesian estimation to convert the measurements of RPM contrast $\Vec{y}$ to an estimate of the qubit population, see Fig.~\ref{fig:RPM-data}b. For this purpose, we update the prior knowledge $P\left(p\right)$ with the likelihood function $ P \left( \Vec{y} | p \right)$, which yields the posterior probability distribution of the residual qubit population
\begin{equation}
    P \left( p | \Vec{y} \right) \propto  P \left( \Vec{y} | p \right) P\left(p\right) .
\end{equation}
The proportionality factor accounts for the normalization of the posterior distribution. We assume that the only prior knowledge of the $\ket{e}$ population is that it is between 0 and 1, so we use a prior probability distribution $P\left(p\right)$ that is flat between 0 and 1.
Averaging the data in Fig.~\ref{fig:RPM-data}a yields a sample mean $\mu_4$ and the standard error of the mean $\sigma_4$. Because the averaged measurements are normally distributed, the probability of observing our data given a true qubit population $p$ forms a Gaussian likelihood function, $P(\Vec{y}|p) = \mathcal{N}(p; \mu_4, \sigma_4)$.
The product of prior and likelihood gives the posterior probability distribution, which is a Gaussian truncated at 0 and 1. The mean of the posterior distribution of qubit population is $\left< p \right> = 8.3 \times 10^{-5}$ (Fig.~\ref{fig:RPM-data}b), with the $95 \% $ confidence interval $\sigma=\left[ 0.027, 2.52 \right] \times 10^{-4}$. The inferred qubit population corresponds to the effective temperature of 26 mK. 



\begin{figure}
  \centering
  \includegraphics[width=1.0\columnwidth]{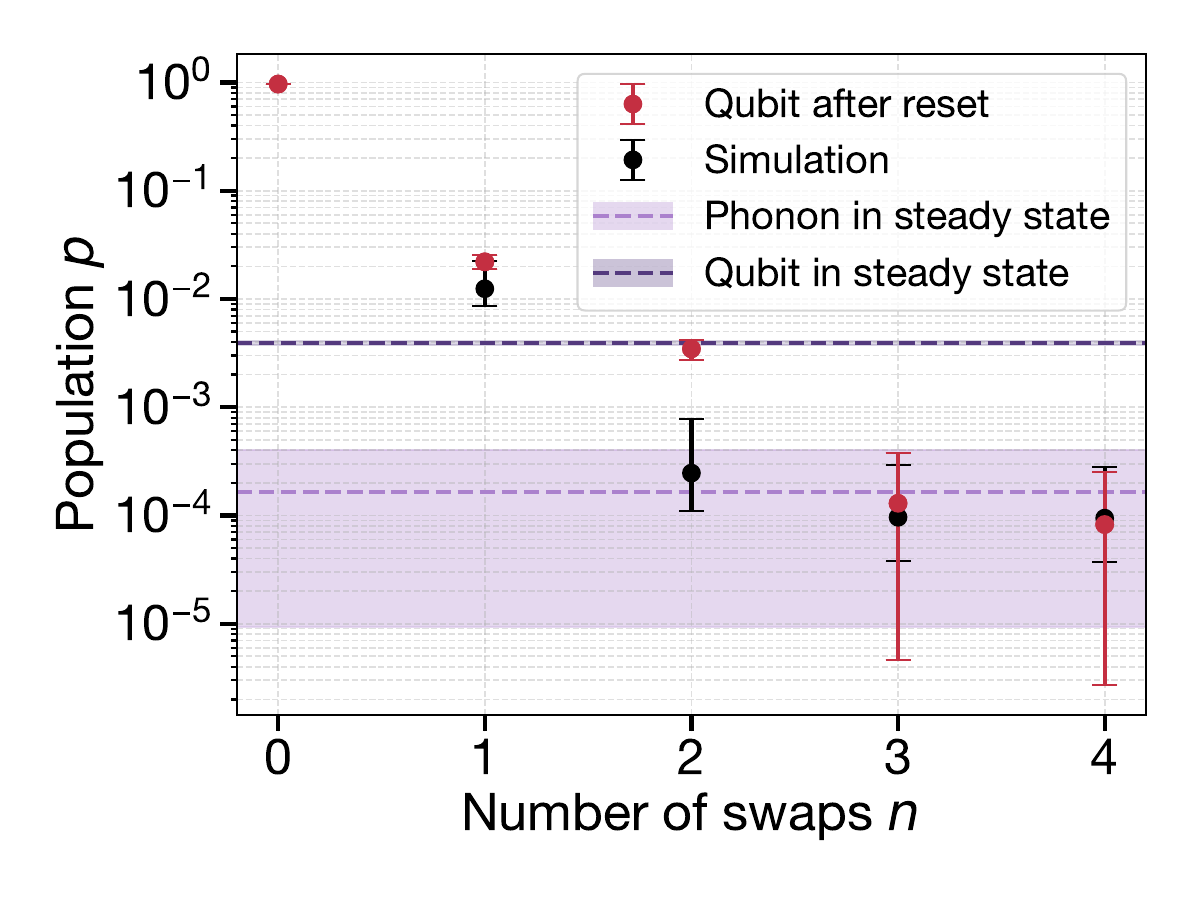}
  \caption{\textbf{Residual qubit population vs. number of iSWAP gates.}
  The measured residual population of the qubit prepared in $\ket{e}$ after a number of qubit-phonon swaps (red) is compared to the measured residual steady-state qubit population of the qubit prepared in $\ket{g}$ without reset (dark purple dashed line). The light purple dashed line indicates the measured steady-state population of the phonon modes, with the shaded region representing the 95\% confidence interval. Black points represent results of a QuTiP master equation simulation, with error bars inherited from the 95\% confidence intervals on the bare system parameters as quoted in the main text.}
  \label{fig:population-vs-swap-number}
\end{figure}

We repeat this characterization for a reset process with a varying number of iSWAP gates and present the results in Fig.~\ref{fig:population-vs-swap-number}. The occupation number with no swaps provides an estimate for the initial $\pi_{ge}$-pulse fidelity of $F_{\pi_{ge}}=96.6\%$. A single swap yields residual $\ket{e}$ populations on the order of $2\%$. Two swaps already cool the transmon to its effective steady-state bath temperature, corresponding to $\ket{e}$ populations on the order of $\sim 0.35\%$.
After three and four swaps, the qubit population saturates at the level of $p_3=1.3 \times 10^{-4}$ and $p_4=8.3 \times 10^{-5}$, respectively. This level corresponds to the steady state Fock 1 state population of an HBAR phonon mode (dashed pink line and shaded confidence interval). This phonon population was independently measured by omitting the initial  
$\pi_{ge}$-pulse while performing the sequence of 4 iSWAP gates (similarly as in  \cite{Omahen2025}). The agreement of $p_3$ with this steady-state confidence interval confirms that the protocol effectively depletes the qubit population to the level of the steady-state phonon population. Moreover, $p_3$ and $p_4$ are consistent with the measured steady-state phonon population in a different device \cite{Omahen2025}, demonstrating the reproducibility of our reset scheme.


We identify the primary limitation of our qubit reset protocol to be the infidelity of the iSWAP gates, which is the core building block of the protocol.
This infidelity predominantly originates from the qubit's effective bath temperature $T_{\text{bath}} = 45 \pm 4~\text{mK}$, and the coherence and dephasing times ($T_1= 23.1 \pm 12.4~\mu\text{s}$, $T_\phi = 17.1 \pm 8.1~\mu\text{s}$). 
In Fig.~\ref{fig:population-vs-swap-number}, we include the results of a master equation simulation that uses the above parameters.
We find good overall agreement between the measured and simulated population, indicating that the main error sources are well identified. We however observe a small discrepancy between the data and simulation for the 2-iSWAP case, which we attribute to qubit frequency fluctuations that may have caused the iSWAP operation in that experiment to be slightly off-resonant.

In addition, another source of error not modeled in the simulation comes from the off-resonant Jaynes-Cummings interaction of the qubit and excited acoustic modes in the dispersive regime \cite{Blais2004} during qubit operations and readout. There, the qubit weakly interacts with the phonon modes that were used in the previous reset steps. Using QuTiP simulations (Appendix \ref{app:hybridization}), we estimate this effect to lead to an excited state population of at most $P_e \approx 1.4 \times 10^{-4}$, consistent with our measurement data. 


Finally, vice versa, single-qubit operations can also unintentionally excite nearby phonon modes. As discussed in Appendix \ref{app:phonon-driving}, this is particularly relevant for the preparation $\pi$-pulse on the qubit that induces an off-resonant displacement of the closest phonon mode. However, both of the aforementioned effects can be strongly suppressed by increasing the qubit-phonon detuning $\Delta$.

\section*{Discussion}

\begin{figure}
  \centering
  \includegraphics[width=1.0\columnwidth]{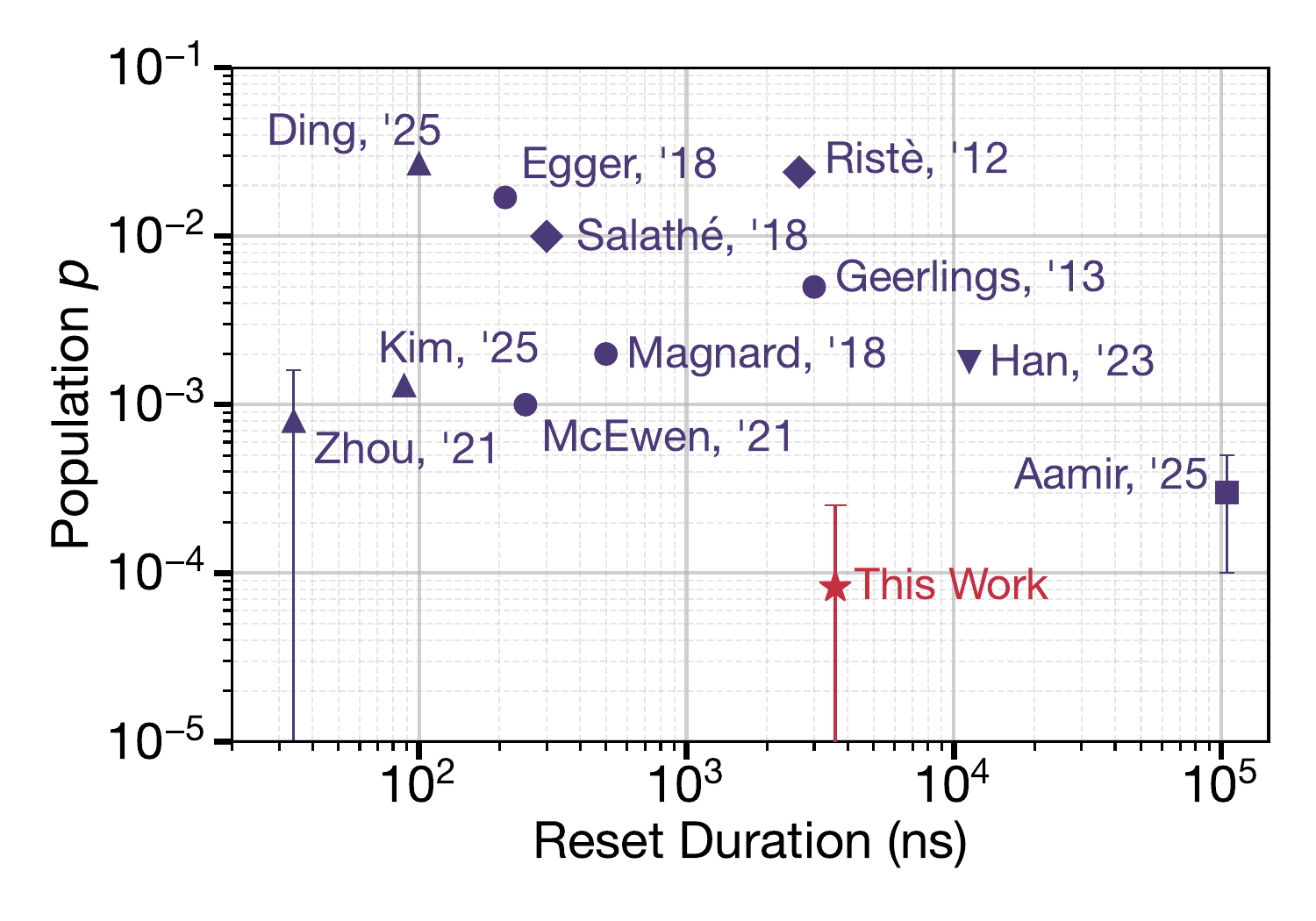}
  \caption{\textbf{Comparison of transmon qubit reset protocol performance.} The plot markers indicate the reset strategy: measurement-based reset schemes (diamonds) \cite{Riste2012, Salathe2018}; reset through engineered dissipation using microwave resonators (disks) \cite{Geerlings2013, Magnard2018, Egger2018, mcewen_removing_2021}; reset using additional active, on-chip microwave elements (upwards triangles) \cite{Ding2025, Zhou2021, kim_fast_2025} and external RF components (triangle down) \cite{Han2023}; reset through microwave bath engineering (square) \cite{Aamir2025}; and reset using a physically different bath (star; this work).}
  \label{fig:comparison}
\end{figure}

Our approach to superconducting qubit reset differs fundamentally from other schemes in that it couples the qubit to a physically distinct bath - a mechanical system rather than an electromagnetic circuit environment typically used in circuit QED \cite{Omahen2025}. This approach enables record-low residual excited-state populations, as illustrated in Fig.~\ref{fig:comparison}. This makes the scheme particularly attractive for applications in which reset fidelity, rather than absolute speed, is the dominant figure of merit - such as noise-limited sensing protocols.

An additional advantage of the HBAR-based approach is the intrinsic multimode structure of the acoustic resonator.
Specifically, measurement-induced excitations into higher states of the transmon qubit is a well-known issue in circuit QED \cite{sank2016a}. We can potentially empty the population in these states by tuning additional qubit transitions into resonance with various near-resonant acoustic modes. For example, the $\ket{f}$-level could be reset by performing an iSWAP gate via the interaction
$\left| f \right>_q \left| 0 \right>_i \leftrightarrow  \left| e \right>_q \left| 1 \right>_i $. 
Leveraging the acoustic spectrum in this way enables repeated extraction of entropy in a hardware-efficient manner.

The primary additional requirement of our approach is the integration of a high-overtone bulk acoustic resonator with a superconducting qubit. While HBARs are not yet a standard component in circuit QED laboratories, they have been successfully coupled to transmon qubits by multiple groups \cite{Chu2018,Crump2023,vonLuepke2024,Potts2025}. Importantly, once integrated, the HBAR acts as a passive auxiliary element and does not require additional microwave components, drive lines, complex pulse shaping, or active feedback. The added fabrication complexity is thus traded for a simplified control architecture.

The speed and fidelity of our reset scheme are limited by the coupling between the HBAR and the qubit. A stronger coupling would increase the speed of the iSWAP operations. Since this shortens the time during which the system can rethermalize with the comparably hot qubit bath, we expect higher reset fidelities. At the same time, the coupling cannot be made arbitrarily strong, as this increases the residual interaction when the qubit is detuned from the phonon modes. This introduces errors in parts of the protocol where the two systems are intended to be fully decoupled, as discussed above and in Appendix~\ref{app:error}. This tradeoff means that we need to repeat the reset sequence for a small number of sequential swaps to reach the target occupation (see Appendix~\ref{app:full_protocol}). Even so, the resulting reset duration of 3.4 $\mu s$ is comparable to widely used reset protocols \cite{Riste2012, Geerlings2013} and occupies a middle ground in the range of reset durations while excelling in terms of fidelity, as shown in Fig.~\ref{fig:comparison}. 
Ultimately, the reset fidelity is limited by the residual population of the phonon modes, which is set by imperfect thermal anchoring, residual heating from external mechanical vibrations, unintended coupling to electromagnetic fields via the piezoelectric transducer, or rare high-energy events such as cosmic rays. 


Beyond qubit initialization, our results establish HBARs as a promising resource in superconducting quantum circuits, complementing recent demonstrations of mechanical resonators as building blocks for universal gate sets \cite{Yang2026} and long-lived quantum memories \cite{luo2025,Garcia-Belles2025}.

\section*{Data availability}
All data are available from the corresponding authors upon reasonable request.

\section*{Acknowledgments}
We thank Yu Yang, Raquel Garcia-Belles and Matteo Fadel for useful discussions.
Fabrication of the quantum devices was performed at the ETH Zurich FIRST cleanroom and the BRNC cleanroom at IBM Zurich. This work was funded by the National Research Council of
Science \& Technology (NST) grant by the Korea government (MSIT) (No. GTL25011-000), the Swiss National Science Foundation (SNSF) under grant $200021\_204073$, and the QuantERA II Program that has received funding from the European Union’s Horizon 2020 research and innovation program under grant agreement no 101017733, and with the SNSF. 

\section*{Author Contributions}
A.O. and S.S. planned and performed the experiment, analyzed the data and derived the results. A.O. and I.K. ran the master equation simulations. A.O., S.S. and Y.C. wrote the manuscript. Y.C. supervised the project.

\appendix
\onecolumngrid

\newpage

\section*{Supplementary Information}

\section{Detailed schematics of the reset protocol}
\label{app:full_protocol}
\begin{figure}[h]
  \centering
  \includegraphics[width=0.8\columnwidth]{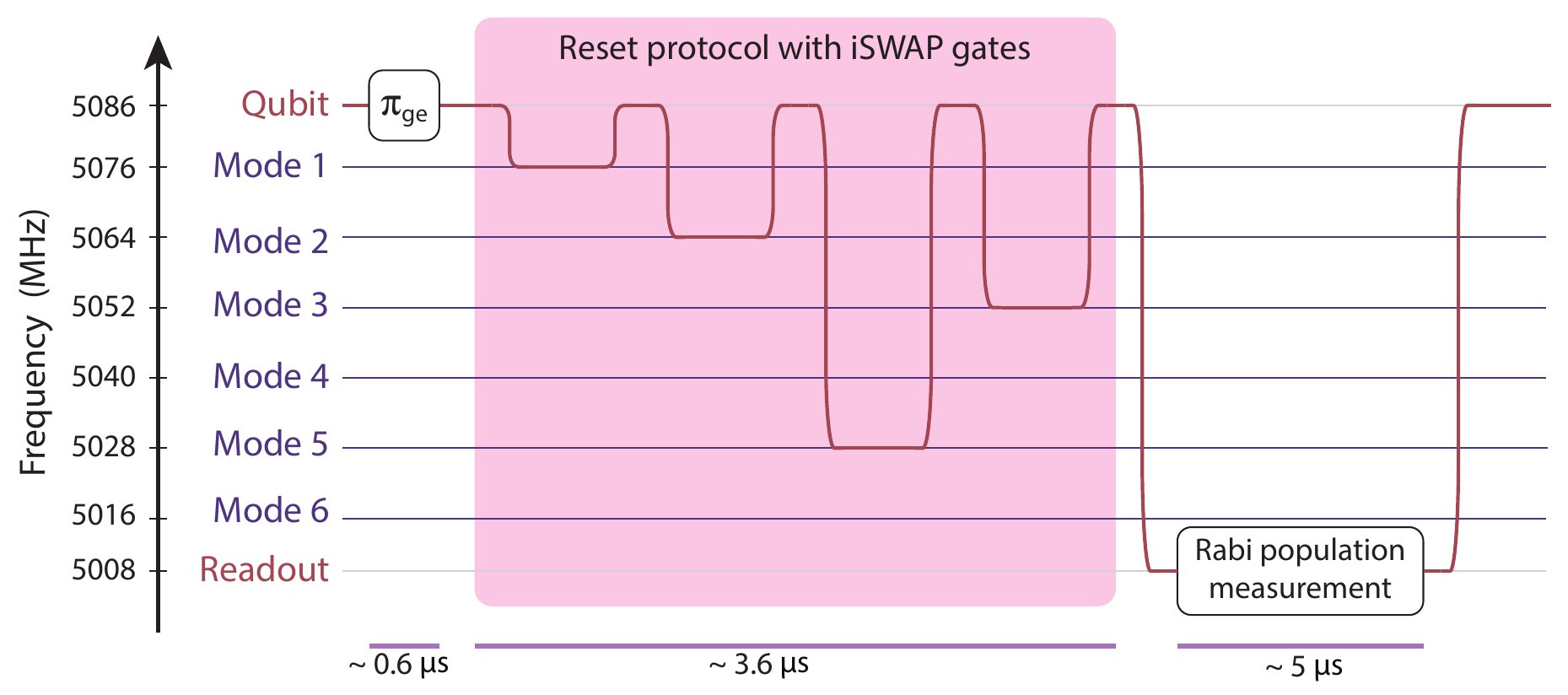}
  \caption{\textbf{Detailed schematics of the reset protocol.}  
  }
  \label{fig:SI-protocol}
\end{figure}

We present the detailed schematics and characterization of the reset protocol in Fig. \ref{fig:SI-protocol}. The sequence begins by exciting the qubit with a $\pi_{ge}$ pulse, followed by a series of iSWAP gates with the phonon modes 1, 2, 5, and 3. Here, the modes are labeled in order of their detuning from the qubit frequency at zero Stark shift and do not necessarily correspond to the labels in the circuit diagram of Fig. \ref{fig:protocol}.
To tune the qubit into resonance with each respective phonon mode, we apply an off-resonant cavity drive to induce an AC Stark shift. In principle, the order of the acoustic modes used in the protocol is flexible. However, we make a few choices to optimize the performance. First, because large Stark shifts accelerate qubit dephasing, we strategically assign the final iSWAP gate to mode 2, which requires a comparatively small frequency shift. 
Furthermore, we exclude mode 4 from the sequence to limit the wait time between measurements. To ensure ground-state relaxation, this waiting time is $6T_1$ of the longest-lived participating mode. Since mode 4 has a lifetime of 400~$\mu$s compared to 60--250~$\mu$s for the others, omitting it increases the overall repetition rate.
Finally, prior to readout, the qubit is detuned to a frequency lower than all utilized modes. Because the readout process dispersively shifts the qubit to even lower frequencies, this placement prevents population in the phonon modes from leaking back into the qubit.

\section{Population of the second excited state}
\label{app:f_level}
Our protocol measures the population \textit{difference} between the $\ket{e}$ state and the second-excited state $\ket{f}$. Therefore, our conclusion that the $\ket{e}$ state has been successfully reset relies on the assumption that $\ket{f}$ state population is negligible compared to the $\ket{e}$ state population. 
In principle, our measurement of $\left< p \right> = 8.3 \times 10^{-5}$ is also consistent with a situation where the $\ket{e}$ and $\ket{f}$ states have much higher populations that are within $8.3 \times 10^{-5}$ of each other. Although this is a highly unlikely scenario, we address it in the following by providing physical arguments against substantial $\ket{f}$-state populations. 

We identify three primary mechanisms for population leakage into the $\ket{f}$-level.
The first potential source of leakage are measurement-induced transitions \cite{sank2016a}, where a microwave readout pulse excites the qubit into higher energy states. In our reset characterization sequence, this effect is mitigated by a delay of $1.4\,\mathrm{ms} \approx 35 \times T_1^q$ at the end of each measurement sequence, during which it is reasonable to assume that the qubit returns to thermal equilibrium before the next iteration. 


A second potential leakage mechanism arises from active control pulses, specifically due to the spectral overlap of the pulses with the $\ket{e}-\ket{f}$ transition. In our protocol, the only active drive is the initial $\pi_{ge}$ pulse used to excite the qubit. To verify that spectral leakage from this pulse is negligible, we performed thermometry measurements on the phonon modes without applying the $\pi_{ge}$ pulse \cite{Omahen2025}. If significant leakage to $\ket{f}$ were present, the sequence including the $\pi_{ge}$ pulse would yield a lower measured population than the sequence without it. However, as shown in Fig.~\ref{fig:population-vs-swap-number}, the measured steady-state phonon population, which involves a single iSWAP with no initial $\pi_{ge}$ pulse, is identical within uncertainties to the population observed after a reset utilizing 3 or 4 iSWAP gates, indicating that the control pulses do not introduce substantial $\ket{f}$-level population.

Finally, we consider the possibility of leakage due to an elevated electromagnetic bath temperature or broadband noise resonant with the $\ket{e}-\ket{f}$ transition, which could come from blackbody radiation from higher cryostat stages or amplifier noise leaking into the sample space. 
We rule this out by analyzing the statistical properties of our data compared to theoretical models. The measured population dynamics in Fig.~\ref{fig:population-vs-swap-number} exhibit excellent agreement with master equation simulations that assume an initial $\ket{f}$-state population of zero ($P_{f}=0$). Moreover, a physical heating mechanism strong enough to maintain a non-equilibrium steady-state population of $P_{f} \sim 1\%$ would likely exhibit temporal fluctuations (e.g., due to $1/f$ noise, amplitude drift, or the inherent statistics of a thermal bath), which would result in observable variance in the qubit population over time. However, our data in Fig. \ref{fig:population-vs-swap-number} show no evidence of such instability, as the populations with 3 and 4 iSWAP gates are consistent with the phonon steady-state population. The observed stability of the measured populations at the $10^{-4}$ level provides strong evidence against the presence of a significant, fluctuating $\ket{f}$-state population.

To provide experimental evidence against a significant $\ket{f}$-state population, we performed additional measurements during a separate cooldown. We note that due to an unintended thermal link in the cryostat during this run, the baseline measured phonon populations were higher for the subsequent measurements. We first performed phonon thermometry without the initial $\pi_{ge}$ pulse (Fig.~\ref{fig:SI-protocol}), yielding a population of $\left< p_{\ket{f} \text{ ref}} \right> = (4.00 \pm 0.37) \times 10^{-3}$ when using the $\ket{f}$ state as the measurement reference. To verify that this signal is not skewed by an elevated $\ket{f}$-state population, we repeated the measurement using the third-excited state, $\ket{h}$, as the reference. This was achieved by applying a $\pi_{fh}$ pulse immediately before the Rabi population measurement to swap the $\ket{f}$ and $\ket{h}$ populations. Using the $\ket{h}$ level as the reference, we measured $\left< p_{\ket{h} \text{ ref}} \right> = (3.72 \pm 0.30) \times 10^{-3}$. The agreement between these measurements within experimental uncertainty indicates that the $\ket{f}$-state population does not significantly exceed the population of the higher-energy $\ket{h}$ state up to the experimental uncertainty, supporting our assumption that it remains negligible. With this additional measurement, the only scenario that we have not ruled out and that would invalidate our main results would be one where the $\ket{e}, \ket{f}, \ket{h}$ states all have elevated populations that are equal to each other within the uncertainty of our thermometry measurements, which is highly unlikely.

In conclusion, we find no evidence for a significant $\ket{f}$-state population that would invalidate our conclusion that the $\ket{e}$ is effectively reset.

\section{Additional error sources}
\label{app:error}
Complementing the discussion in the main text, here we analyze two additional error sources that could potentially limit the fidelity of the reset protocol. Specifically, we discuss the off-resonant hybridization of the qubit and acoustic modes, as well as the excitation of the acoustic modes during qubit control pulses.

\subsection{Off-resonant Jaynes-Cummings interaction}
\label{app:hybridization}
In our reset protocol, we transfer the majority of the qubit population into phonon mode 1, which is used in the first iSWAP operation. Throughout the rest of the protocol, the mode 1 population can be transferred back to the qubit through their off-resonant Jaynes-Cummings interaction. In the simplified picture of a qubit dispersively coupled to a single acoustic mode $i$, the generalized Rabi frequency is $\Omega= \sqrt{\Delta^2+4g^2}\approx \Delta$, where $\Delta=\omega_q - \omega_i$. The population exchange is therefore suppressed by a factor $(g/\Omega)^2$. 

The effect of unwanted off-resonant excitation transfer is the largest at the end of the protocol, during the final iSWAP with mode 3. This is because no further iSWAPs are performed afterwards to provide further cooling, and the effect of off-resonant interaction during readout is smaller due to larger detuning. To estimate this effect, we perform QuTiP simulations with the Hamiltonian 
\begin{equation}
    H = \frac{\Delta}{2} \sigma_z + \Delta a_3^\dagger a_3 + g (\sigma_+ a_1 + \sigma_- a_1^\dagger) + g (\sigma_+ a_3 + \sigma_- a_3^\dagger),
\end{equation}
where $\sigma_z$ is the Pauli-Z operator for the qubit, and $\sigma_{\pm}$ are the qubit raising and lowering operators. The operators $a_i^\dagger$ and $a_i$ denote the creation and annihilation operators for mode $i \in \{1, 3\}$. The detuning during the last iSWAP is $\Delta \approx 2\pi \cdot 25$ MHz (Fig. \ref{fig:SI-protocol}), while the coupling strength $g\approx 2\pi \cdot 300$ kHz. Here we do not include decoherence, since the lifetime of the phonon modes are relatively long. In the extreme scenario where the system begins in state $\ket{g}_q \ket{1}_1 \ket{0}_3$ before the final iSWAP gate, the simulation suggests an upper bound on the final qubit population of $P\approx 1.4 \times 10^{-4}$, consistent with our measurements.

\subsection{Acoustic mode excitation during qubit control pulses}
\label{app:phonon-driving}

A second potential limitation arises from the spurious excitation of the acoustic modes during qubit control pulses. To quantify this, we consider the Hamiltonian of a qubit coupled to a single acoustic resonator mode in the laboratory frame
\begin{equation}
    H_{\text{lab}} = \underbrace{\frac{\omega_q}{2}\hat{\sigma}_z + \omega_r \hat{a}^\dagger \hat{a}}_{H_0} + \underbrace{g(\hat{a}\hat{\sigma}_+ + \hat{a}^\dagger \hat{\sigma}_-)}_{V} + \underbrace{2\Omega \cos(\omega_d t)(\hat{\sigma}_+ + \hat{\sigma}_-)}_{H_{\text{drive}}(t)} ,
\end{equation}
where $\omega_q$ and $\omega_r$ are the qubit and acoustic mode frequencies, respectively. The lowering operator of the qubit is denoted by $\hat{\sigma}_-$, and $\hat{a}$ is the annihilation operator for the acoustic mode. The qubit drive pulse is characterized by amplitude $\Omega$ and frequency $\omega_d$. 

We then proceed by moving into a frame rotating at the drive frequency $\omega_d$ via the unitary transformation $U(t) = \exp\left[ i \omega_d t \left( \hat{\sigma}_z/2 + \hat{a}^\dagger \hat{a} \right) \right]$. After applying the rotating wave approximation (RWA) and neglecting terms oscillating at $2\omega_d$, we arrive at
\begin{equation}
    H_{\text{RWA}} = \frac{\Delta_q}{2}\hat{\sigma}_z + \Delta_r \hat{a}^\dagger \hat{a} + g(\hat{a}\hat{\sigma}_+ + \hat{a}^\dagger \hat{\sigma}_-) + \Omega(\hat{\sigma}_+ + \hat{\sigma}_-) , 
    \label{eq:HRWA}
\end{equation}
where $\Delta_q = \omega_q - \omega_d$ and $\Delta_r = \omega_r - \omega_d$ are the qubit and acoustic mode detunings relative to the drive.

We operate in the dispersive regime ($|\Delta| = |\omega_q - \omega_r| \gg g$). To decouple the qubit and phonon degrees of freedom to first order in $g/\Delta$, we perform a Schrieffer-Wolff transformation with the generator $S = \frac{g}{\Delta}(\hat{a}\hat{\sigma}_+ - \hat{a}^\dagger \hat{\sigma}_-)$. The resulting effective Hamiltonian in the dressed state basis is then
\begin{equation}
    H_{\text{eff}} = \frac{\tilde{\Delta}_q}{2}\hat{\sigma}_z + \Delta_r \hat{a}^\dagger \hat{a} + \chi \hat{a}^\dagger \hat{a} \hat{\sigma}_z + \Omega(\hat{\sigma}_+ + \hat{\sigma}_-) + \frac{g\Omega}{\Delta}(\hat{a}+\hat{a}^\dagger)\hat{\sigma}_z,
    \label{eq:supp-H-eff}
\end{equation}
where $\tilde{\Delta}_q = \Delta_q + \chi$ is the Lamb-shifted qubit detuning and $\chi = g^2/\Delta$ is the dispersive shift. 

When the control pulse is resonant with the qubit transition ($\tilde{\Delta}_q = 0$), the Hamiltonian simplifies to
\begin{equation}
    H_{\text{eff}} = \Delta \hat{a}^\dagger \hat{a} + \chi \hat{a}^\dagger \hat{a} \hat{\sigma}_z + \Omega(\hat{\sigma}_+ + \hat{\sigma}_-) + \frac{g\Omega}{\Delta}(\hat{a}+\hat{a}^\dagger)\hat{\sigma}_z.
    \label{eq:supp-H-eff-2}
\end{equation}
The first term describes the detuning of the phonon mode from the drive, the second represents the dispersive interaction between qubit and phonon, and the third represents the intended qubit drive. The final term describes a qubit-state dependent displacement of the phonon mode, originating from the qubit drive, and resulting in unwanted phonon mode excitations.

The effective force driving this spurious displacement is proportional to $g\Omega / \Delta$. Since the drive is detuned from the resonator by $\Delta$, the resulting amplitude of the resonator response scales as $g\Omega / \Delta^2$. 
Because the reset protocol subsequently relies on iSWAP gates to exchange excitations between the qubit and the acoustic modes, any spurious population generated in the resonator during control pulses (in our case the initial $\pi_{ge}$ pulse) will be swapped back to the qubit. 

In our sequence, the final iSWAP utilizes mode 3 (Fig.~\ref{fig:SI-protocol}), which is detuned by $\Delta/2\pi = 34$~MHz from the qubit during the initial $\pi_{ge}$ pulse. For the parameters used in our experiment, the amplitude of the resonator response is approximately $g\Omega / \Delta^2 = 4 \times 10^{-4}$. Crucially, the residual population swapped into the qubit scales as the square of this response,  $\left( g\Omega / \Delta^2 \right)^2$, yielding a negligible unwanted population of roughly $1.6 \times 10^{-7}$. 
We note that this value serves primarily as an order-of-magnitude estimate. Because the effective driving force depends on the state of the qubit (Eq.~\ref{eq:supp-H-eff-2}), evaluating the exact system dynamics and residual population requires full numerical simulations. 

While in our case this off-resonant excitation error is heavily suppressed compared to the hybridization error discussed in the previous section, its scaling with the square of the drive amplitude ($\Omega^2$) dictates that it can become a significant error source for fast, strongly driven gates. In that case, this effect should be mitigated either by increasing the detuning $\Delta$, or by employing pulse engineering techniques that ensure the driven phononic state completes a closed loop in phase space.

\newpage

\bibliographystyle{apsrev4-2} 
\bibliography{references_latex_safe_refreshed} 

\end{document}